\documentclass[manuscript]{acmart}

\AtBeginDocument{%
  \providecommand\BibTeX{{%
    \normalfont B\kern-0.5em{\scshape i\kern-0.25em b}\kern-0.8em\TeX}}}

\setcopyright{acmcopyright}
\copyrightyear{2026}
\acmYear{2026}

\acmJournal{JACM}
\acmVolume{37}
\acmNumber{4}
\acmArticle{111}
\acmMonth{8}

\acmPrice{15.00}
\acmISBN{978-1-4503-XXXX-X/18/06}

\usepackage{lettrine}

 \usepackage{color,soul,bm}

\begin{document}

\title{(Un)fair devices: Moving beyond AI accuracy in personal sensing}

\author{Sofia Yfantidou}
\authornote{Work unrelated to current affiliation. The majority of this work was completed while the author was affiliated with Nokia Bell Labs and the Aristotle University of Thessaloniki, Greece.}
\email{sofia@atride.eu}
\orcid{0000-0002-5629-3493}
\affiliation{%
  \institution{ATRIDE}
  \city{Athens}
  \country{Greece}
}

\author{Marios Constantinides}
\authornote{Work partially done at Nokia Bell Labs, United Kingdom.}
\email{marios.constantinides@cyens.org.cy}
\orcid{0000-0003-1454-0641}
\affiliation{%
  \institution{CYENS Centre of Excellence}
  \city{Nicosia}
  \country{Cyprus}
}

\author{Dimitris Spathis}
\email{spathis@google.com}
\orcid{0000-0001-9761-951X}
\affiliation{%
  \institution{Nokia Bell Labs}
  \city{Cambridge}
  \country{UK}
}
\affiliation{%
  \institution{University of Cambridge}
  \city{Cambridge}
  \country{UK}
}
\authornote{Work unrelated to current affiliation at Google. The majority of this work was completed while the author was affiliated with Nokia Bell Labs.}
\affiliation{%
  \institution{Google Research}
  \city{London}
  \country{UK}
}

\author{Athena Vakali}
\email{avakali@csd.auth.gr}
\orcid{0000-0002-0666-6984}
\affiliation{%
  \institution{Aristotle University of Thessaloniki}
  \city{Thessaloniki}
  \country{Greece}
}

\author{Daniele Quercia}
\email{daniele.quercia@nokia-bell-labs.com}
\orcid{0000-0001-9461-5804}
\affiliation{%
  \institution{Nokia Bell Labs}
  \city{Cambridge}
  \country{UK}
}
\affiliation{%
  \institution{Politecnico di Torino}
  \city{Turin}
  \country{Italy}
}

\author{Fahim Kawsar}
\email{fahim.kawsar@glasgow.ac.uk}
\orcid{0000-0001-5057-9557}
\affiliation{%
  \institution{University of Glasgow}
  \city{Glasgow}
  \country{Scotland}
}

\newcommand{\rev}[1]{\textcolor[rgb]{0.00,0.00,0.00}{#1}}
\renewcommand{\shortauthors}{Yfantidou, et al.}

\begin{abstract}
 Personal devices are omnipresent in our lives, seamlessly monitoring our activities, from smart rings tracking sleep patterns to smartwatches keeping an eye on missed heartbeats. The rich data streams from such devices fuel advanced Artificial Intelligence (AI) applications. Instead of solely relying on direct sensor measurements, these applications are increasingly leveraging Machine Learning (ML) model estimates to derive insights. But are these estimates biased or not? This \rev{literature review} delivers compelling evidence about the impact of hidden biases that creep into ML models deployed on personal devices.
 We discuss critical bias issues \rev{drawn from prior work} such as racial bias in pulse oximeters, weight bias in optical heart rate sensors, and sex bias in audio-based diagnostics. In response to these challenges, we advocate for a shift from prioritizing performance-oriented evaluations of personal devices to adopting assessments grounded in a human-centered approach. To facilitate this transition, we provide guidelines for the design, development, evaluation, and use of unbiased AI in personal devices, recognizing their potential impact on improving our health, lifestyle, and productivity---more than any other technology.
 
\end{abstract}



\maketitle
\section{Introduction}\label{introduction}
As an extension of ourselves, omnipresent and intimately close, personal devices continuously sense and learn from physiological and behavioral signals of our physical and mental well-being. Their ubiquity is unmistakable, with nearly one-third of adults in the United States using wearable devices to keep tabs on their health and fitness, while globally, the number of connected wearable devices has more than doubled in \rev{recent} years \cite{forecast2019cisco}. Beyond their prevalence, personal devices have shown versatility across various domains, from workplace monitoring to educational contexts. 
Nevertheless, their most profound impact has been within the healthcare sector, effecting a paradigm shift from reactive to predictive health, with the promise of proactive interventions on the horizon. From detecting arrhythmias through tracking the subtle electrical impulses controlling our hearts to monitoring emotional arousal through the measurement of electrodermal responses, personal devices have evolved into miniature medical instruments, providing a comprehensive understanding of our health right from our pockets. 

\begin{figure}
  \centering
  \includegraphics[width=\linewidth]{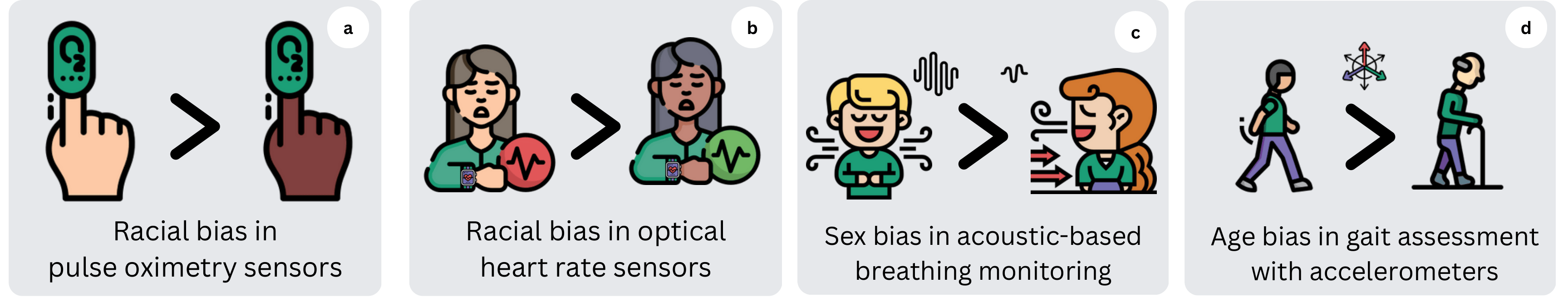}
  \caption{\textbf{\rev{Motivating examples} of biased personal devices \rev{drawn from prior work}.} Research efforts have surfaced \textit{racial biases} in (a) pulse oximetry \cite{fawzy2022racial} and (b) optical heart rate sensors \cite{bent2020investigating}, (c) \textit{sex bias} in acoustic signals \cite{10.1145/3534595}, and (d) \textit{age bias} in accelerometer and gyroscope measurements \cite{10.1145/3351281}.}
  \label{fig:examples}
  \vspace{-0.2in}
\end{figure}
But here \rev{is} the stark reality--we cannot assume that personal devices, with their powerful Machine Learning (ML) algorithms, operate equitably for all users. In this \rev{literature review}, we provide compelling evidence that the data and models fueling these advancements are not immune to biases, possibly rendering algorithmic decisions ``unfair''.  There are two key notions of fairness: individual fairness and group fairness. Individual fairness ensures similar treatment for similar individuals, irrespective of their demographic characteristics. Conversely, group fairness, the focus of this work, emphasizes equitable outcomes across predefined demographic groups.
Real-world cases of ``unfair'' ML algorithms in personal devices abound (Figure~\ref{fig:examples}). 
For example, neural network algorithms trained to perform skin lesion classification showed approximately half the original diagnostic accuracy on black patients \cite{kamulegeya2019using}. At the same time, people of color are consistently misclassified by health sensors such as oximeters as they were scientifically tested on predominantly white populations~\cite{fawzy2022racial}. Yet, the implications of such biases extend beyond mere measurement errors. 
Biases in pulse oximetry sensors during the COVID-19 pandemic led to significant delays in recognizing hypoxia among Black and Hispanic patients, possibly contributing to ethnic disparities in COVID-19 outcomes \cite{holmes2020black}. As technology research prioritizes performance metrics and rapid progress, people become marginalized in the development process.

To systematically corroborate the extent to which research in mobile, wearable, and ubiquitous computing addresses biases in personal devices and to uncover sources of harm within their ML pipelines, we surveyed prominent venues publishing work on systems, human-computer interaction, and ML on personal devices. Surprisingly, \rev{according to our findings}, the percentage of papers reporting fairness assessments ranged from none to a mere 10\% \rev{published between 2018 and 2025}.
However, the lack of awareness transcends academia. 
A recent white paper on validating the Apple watch estimations of cardio fitness with maximal oxygen uptake omitted any reference to race as a baseline characteristic
and refrained from providing performance breakdowns conditioned on demographics, adhering solely to aggregate evaluations \cite{apple2021}. 

Yet, it is time we acknowledged that personal devices are not immune to biases. These devices have evolved beyond mere step counting, now incorporating functionalities ranging from activity recognition and sleep inference to high-stakes applications--detecting Atrial Fibrillation (AFib) \cite{10.1145/3397313}, diagnosing COVID-19 infections \cite{10.1145/3394486.3412865}, and predicting fertility~\cite{maijala2019nocturnal}. Contemplate the ethical and societal repercussions though if an AFib detection algorithm exhibited racial bias or a fertility prediction algorithm proved less reliable for women in non-Western countries. While such applications remain largely unregulated, they fall within the scope of forthcoming regulatory initiatives, such as the EU AI Act, which would mandate comprehensive risk assessments. 

\begin{figure}[t!]
  \centering
  \includegraphics[width=0.72\linewidth]{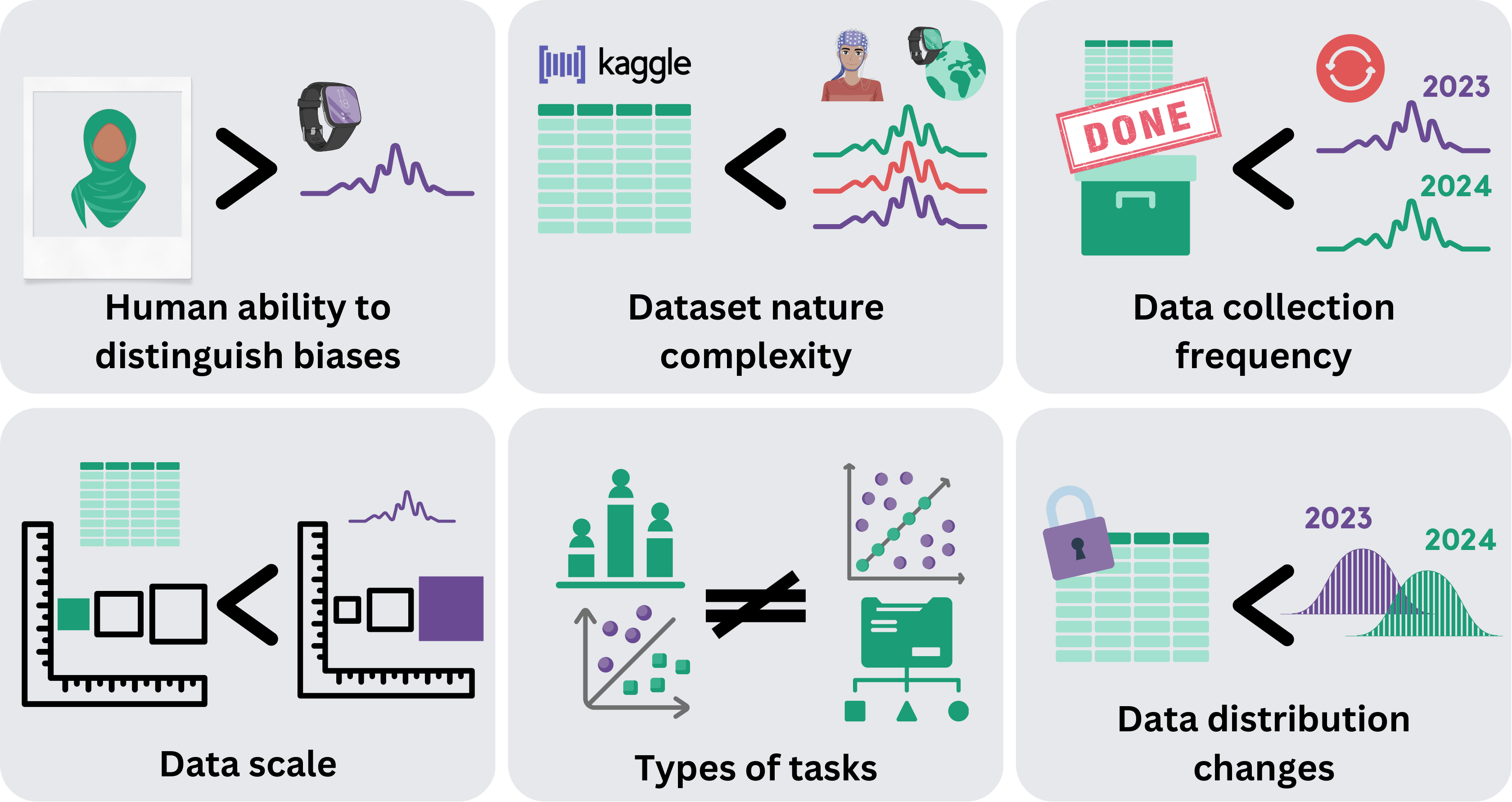}
  \caption{\textbf{Conceptual differences between ML fairness and personal devices data.} Biases are harder to surface, given the complexity of temporal data. Continuous data collection can introduce data drifts, resulting in larger and more dynamic datasets. Additionally, the ML tasks addressed by the personal devices community diverge from those typically explored in fairness research.\label{fig:teaser}} 
  \vspace{-0.2in}
\end{figure}
While fairness in ubiquitous computing has been acknowledged as a challenge since the 1990s \cite{weiser1993some}, research on fairness in personal devices has received comparatively less attention than in the field of ML. Important works in ML offer foundational insights encompassing broad principles applicable across computer science subdomains \cite{zook2017ten}, as well as specific solutions tied to different stages of the ML pipeline, such as 'datasheets for datasets' or 'model cards for model reporting' \cite{mitchell2019model,gebru2021datasheets}. Additionally, researchers focus on raising domain-specific awareness by identifying challenges and offering recommendations, such as addressing bias on the web \cite{baeza2018bias}, in computer vision \cite{buolamwini2018gender}, and in language models \cite{bender2021dangers}.

Yet, personal devices' data and models possess unique characteristics not commonly shared with the broader discourse on AI Ethics (Figure~\ref{fig:teaser}). For example, the mobile and wearable community typically deals with small-scale datasets often collected by the authors in-the-lab or in-the-wild during feasibility studies, while the broader ML community frequently utilizes popular, medium- to large-scale benchmark datasets such as UCI Adult \cite{kohavi1996uci}, German Credit \cite{dua2017uci}, and COMPAS \cite{compas}. Such data are collected once and are immutable, in contrast to personal devices' data that are mutable and longitudinally collected. Contrary to the tabular format of benchmark datasets, such data are mostly sequential/temporal, e.g., sensor data. Then, the biases are harder to surface. In other words, while it is relatively straightforward to distinguish a person's skin tone from an image, it is much harder to do so from oximetry measurements. Similarly, an image captures the entire semantic meaning of a given data point, whereas a sensor signal is just an observation of the effect we measure; a variation in heart rate cannot be explained by the signal alone as it might be caused by external factors (alcohol, stress). As the vision of ubiquitous computing strives to blend technologies in the background, biases are blended, too. However, it is possible to build devices that are both accurate and fair. As research evolves, the mobile and wearable computing community needs to recognize that the initial step, awareness, often takes a backseat in our discussions.

\section{Background and Related Work\label{related-work}}
First, we present the reader with an introduction to fairness: definitions, measurements, and enhancement mechanisms. Next, we situate our review in previous fairness literature covering three broad areas: \emph{a)} comprehensive surveys; \emph{b)} domain-targeted surveys; and \emph{c)} and surveys focused on over-represented populations. 
\smallskip

\noindent\textbf{Fairness Definitions and Measurements.} Fairness entails the ``protection of individuals and groups from discrimination or mistreatment with a focus on prohibiting behaviors, biases and basing decisions on certain protected factors or social group categories''~\cite{fairnessberkeley}. Quantitative fields (e.g., computer science, statistics) view fairness as a mathematical problem of ``equal or equitable allocation, representation, or error rates, for a particular task or problem''~\cite{narayanan21fairness}. 
Fairness enhancement mechanisms fall under three categories \cite{pessach2022review,caton2020fairness}: a) \textit{pre-processing}; b) \textit{in-processing}; and c) \textit{post-processing} mechanisms. Pre-processing mechanisms involve altering the training data before feeding it into an ML model, while in-processing mechanisms involve modifying the ML algorithms to account for fairness during training. Ultimately, post-processing mechanisms involve altering the output probabilities of an ML model to mitigate biases/imbalances (e.g., enforcing a quota). However, due to the late stage in the learning process in which they are applied, post-processing mechanisms commonly obtain inferior results \cite{woodworth2017learning}. They are also considered too invasive or discriminatory since they deliberately damage accuracy for some subjects to compensate others \cite{pessach2022review}; hence they are less frequently preferred in practice.
\smallskip

\noindent\textbf{Comprehensive Surveys.} Addressing algorithmic bias in ML has been a longstanding issue \cite{caton2020fairness}, despite its recent surge. A number of comprehensive surveys shed light on data and model biases across domains and compared potential mitigation solutions. For example, \citet{caton2020fairness} and \citet{pessach2022review} discussed fairness metrics and categorized mitigation approaches into a widely accepted framework of pre-processing, in-processing, and post-processing methods independently of the application domain. \citet{wan2022processing} focused exclusively on in-processing modeling methods such as adversarial debiasing, disentangled representations, and fairness-aware data augmentation, while  \citet{pessach2022review} provided an overview of emerging research trends, including fair adversarial learning, word embeddings, and recommender systems. Along these lines, \citet{le2022survey} surveyed available datasets for fairness research, including financial, criminological, healthcare, social, and educational datasets. Yet, despite these surveys' considerable contributions, they tend to be of generic nature and rarely discuss data, models, and applications related to an individual community. 
\smallskip

\noindent\textbf{Domain-targeted Surveys.} Another line of work took a deep dive into well-defined domains, for example, by focusing on fairness for ML for graphs~\cite{choudhary2022survey}, on exploring notions of fairness in clustering~\cite{chhabra2021overview}, and on studying fairness in recommender systems~\cite{li2022fairness}. Another group of works targeted specific unprivileged groups or high-stakes domains. For example, \citet{olteanu2019social} reviewed the literature surrounding social data biases, such as biases in user-generated content, expressed or implicit relations between people, and behavioral traces, while in \cite{sun2019mitigating}, the authors focused specifically on gender bias in Natural Language Processing (NLP). On a different note, \citet{10.1145/3173574.3174156} featured emerging trends for explainable, accountable, and intelligible systems within the human-computer interaction community, also discussing notions of fairness. Closer to our work, \citet{mhasawade2021machine} discussed ML fairness in the domain of public and population health, and \citet{xu2022algorithmic} explored algorithmic fairness in computational medicine, which only covers a subset of the broad, interdisciplinary mobile, wearable, and ubiquitous computing research domain.
\smallskip

\noindent\textbf{Surveys Focused on Over-represented Populations (WEIRD Research).} 
WEIRD research refers to a common criticism in the social sciences that much of the research is conducted on a sample of participants that is \textbf{W}estern, \textbf{E}ducated, \textbf{I}ndustrialized, \textbf{R}ich, and \textbf{D}emocratic. In particular, a comprehensive study conducted by \citet{henrich2010weirdest} in 2010 revealed a significant bias in sample populations, and found that most research samples come from WEIRD populations, which represent only 12\% of the global population but account for 96\% of research samples. 
Similarly, \citet{linxen2021weird} conducted a meta-study on CHI proceedings from 2016 to 2020, reporting that 73\% of those papers are based on Western populations. Another recent meta-study on FAccT proceedings from 2018 to 2021 found highlighted concerns with biases in word embeddings and computer vision, and racial disparities, while only a $\sim10\%$ of those papers used original, empirical datasets \cite{laufer2022four}. Similarly, by analyzing FAccT proceedings between 2018 and 2022, Septiandri et al.~\cite{ali_weird_2023} found that 84\% of the analyzed papers were exclusively based on participants from Western countries, particularly exclusively from the U.S. (63\%).
\smallskip

It is evident that research communities other than mobile, wearable, and ubiquitous computing  have already started to explore ways of reporting data and models in a fair manner to surface and, ultimately, address encountered biases. Yet, the state of fairness in this community remains unknown, as, at the time of writing, there exists no other survey or position paper in the intersection between personal devices and fairness.

\section{Methodology}
\label{sec:methodology}
Next, we delineate our methodology for conducting this literature synthesis (\S\ref{conducting}) and provide our positionality statement (\S\ref{positionality}).

\begin{figure}[tb!]
  \centering
  \includegraphics[width=\linewidth]{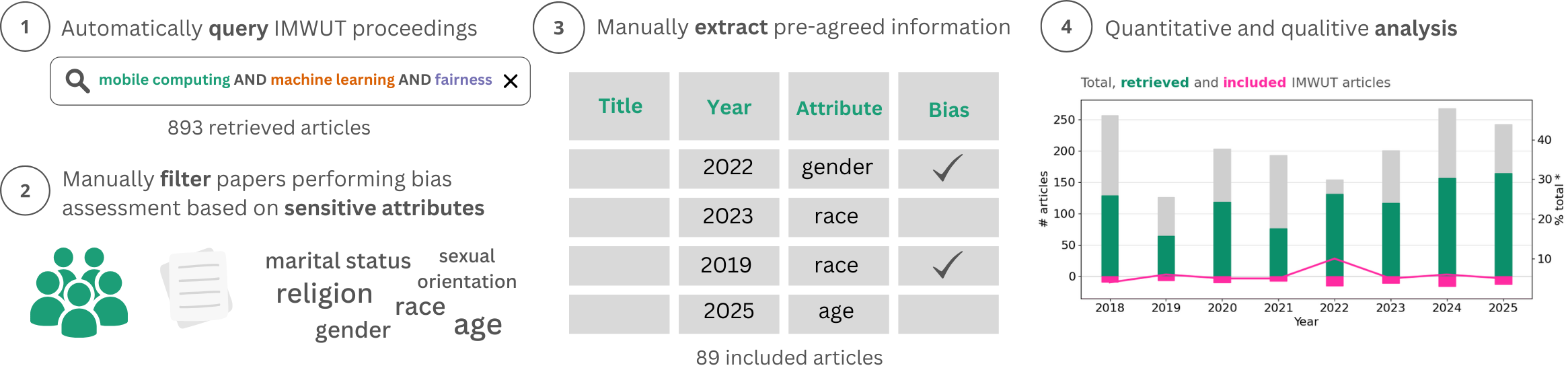}
  \caption{\textbf{Illustration of the literature synthesis methodology.} A high-level overview of the process, including the querying, manual extraction, and filtering, and one of the main results. Note that for readability purposes, we present a simplified version of our query and data extraction. Our query retrieves \rev{$\sim58\%$} of all IMWUT publications, while our eligibility assessment filtering ends up with \rev{89} papers ($\sim9\%$ of retrieved papers). We notice that only a very small fraction of all IMWUT papers looks at fairness issues, with only a small deviation across years.\label{fig:methodology}}
\end{figure}
\subsection{Conducting the Literature Synethesis\label{conducting}}
We followed Kitchenaham and Charters' protocol \cite{kitchenham2007guidelines} for conducting systematic reviews to ensure the quality of included works and limit the initial retrieved papers. At least three authors were involved \rev{during the eligibility assessment} to minimize the effects of bias and priming. In accordance with this protocol, we initially identified the need for a systematic review, as discussed in Section~\ref{introduction}, namely to explore where personal device fairness overlaps with traditional fairness definitions, in what ways and to which extend. Figure~\ref{fig:methodology} provides a high-level overview of the process.
\smallskip 

\begin{figure}[tb!]
  \centering
  \includegraphics[page=1,width=.75\linewidth]{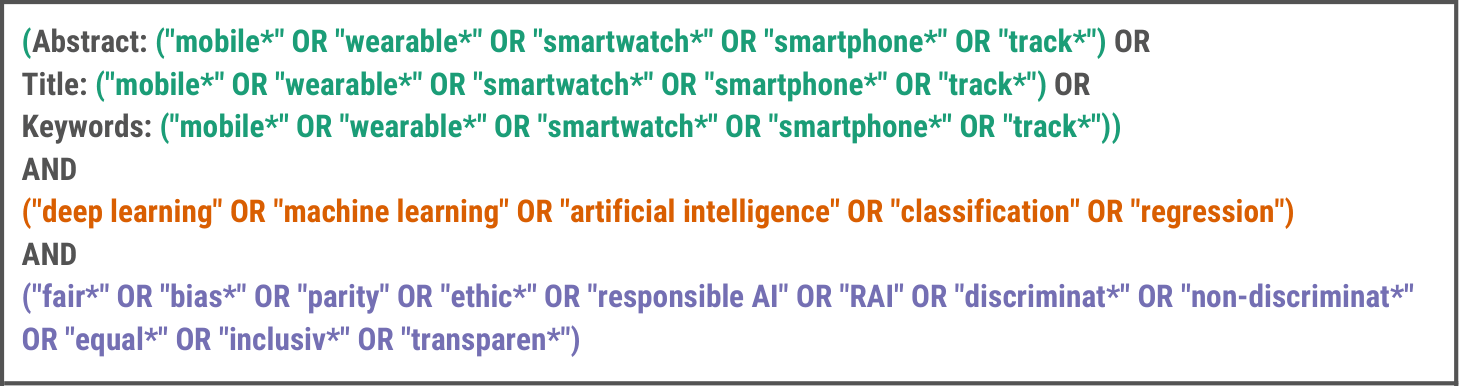}
  \caption{\textbf{The query utilized for recovering relevant papers from the ACM Digital Library}. Terms related to mobile, wearable, and ubiquitous computing are highlighted in green, ML in orange, and fairness in purple.\label{fig:query}}
\end{figure}

\noindent\textbf{Paper Identification \& Screening.} 
Mobile, wearable, and ubiquitous computing research crosses several fields, ranging from Human-Computer Interaction, Hardware and Software Systems, and Knowledge Discovery and Data Mining. For the scope of this review, we focused on the Proceedings of the ACM on Interactive, Mobile, Wearable and Ubiquitous Technologies (IMWUT), a premier journal in the community.
For the search process, we utilized the ACM Digital Library, focusing on papers that were published between 2018 and \rev{2025} to capture emerging trends in fairness and personal device research. Apart from year filtering, for the most part, we did not limit our search to meta-data, such as titles, keywords, and abstracts, but rather we expanded it to any searchable field, including full text. That excludes the first part of the query, which tries to match terms such as wearable(s) or mobile(s) only in the papers' meta-data, as seen in Figure~\ref{fig:query}. 
\smallskip

\noindent\textbf{Query Definition.} For the definition of our query, we followed similar terminology with relevant review papers in the fairness literature \cite{caton2020fairness}\rev{\footnote{The authors use ``fairness datasets'' as the primary query term along with other terms like ``bias,'' and ``discrimination,'' to narrow down the search.}}. Additionally, according to Fjeld et al.'s analysis of prominent AI principles documents, \cite{fjeld2020principled}, ``the fairness and non-discrimination theme is the most highly represented theme in our dataset, with every document referencing at least one of its six principles: ``non-discrimination and the prevention of bias'', ``representative and high-quality data'', ``fairness'', ``equality'', ``inclusiveness in impact'', and ``inclusiveness in design'', mostly included in our query's coverage. To capture the industrial perspective, we consulted the Responsible Artificial Intelligence (RAI) white papers issued by large tech companies. Specifically, Google's\footnote{\url{https://ai.google/responsibilities/responsible-ai-practices/?category=fairness}} and Meta's\footnote{\url{https://ai.facebook.com/blog/facebooks-five-pillars-of-responsible-ai/}} RAI principles talk about ``fairness and inclusion'', Amazon's\footnote{\url{https://aws.amazon.com/machine-learning/responsible-machine-learning/}} RAI principles promote ``diversity, equity, and inclusion'' through ``detecting bias''. Similarly, Nokia's\footnote{\url{https://www.bell-labs.com/institute/blog/introducing-nokias-6-pillars-of-responsible-ai/}} RAI fairness pillar talks about ``fairness, non-discrimination, accessibility, and inclusivity''. Thus, an iterative refinement process resulted in the query shown in Figure~\ref{fig:query}. \rev{Notably, we omit the term ``personal device'' from the final query, as an ablation check\footnote{By ablation, we refer to a controlled query-variation analysis in which additional terms are added to the baseline search query to assess their contribution. Specifically, we augmented the device-related component of the query (Figure~\ref{fig:query}, green section) with the terms ``personal device*'' and ``personal sensing'' and re-ran the search for the original time range (2018--2022). In both cases, the query returned 523 publications, identical to the baseline query.} showed that adding the terms ``personal device*'' or ``personal sensing'' within the inspected time range did not retrieve any additional publications, indicating that the ubiquitous and wearable computing terminology already provides complete coverage of the relevant literature.}

\begin{figure}[tb!]
  \centering
  \includegraphics[
    width=.9\linewidth,
    trim=0 1.5cm 0 3cm,
    clip
  ]{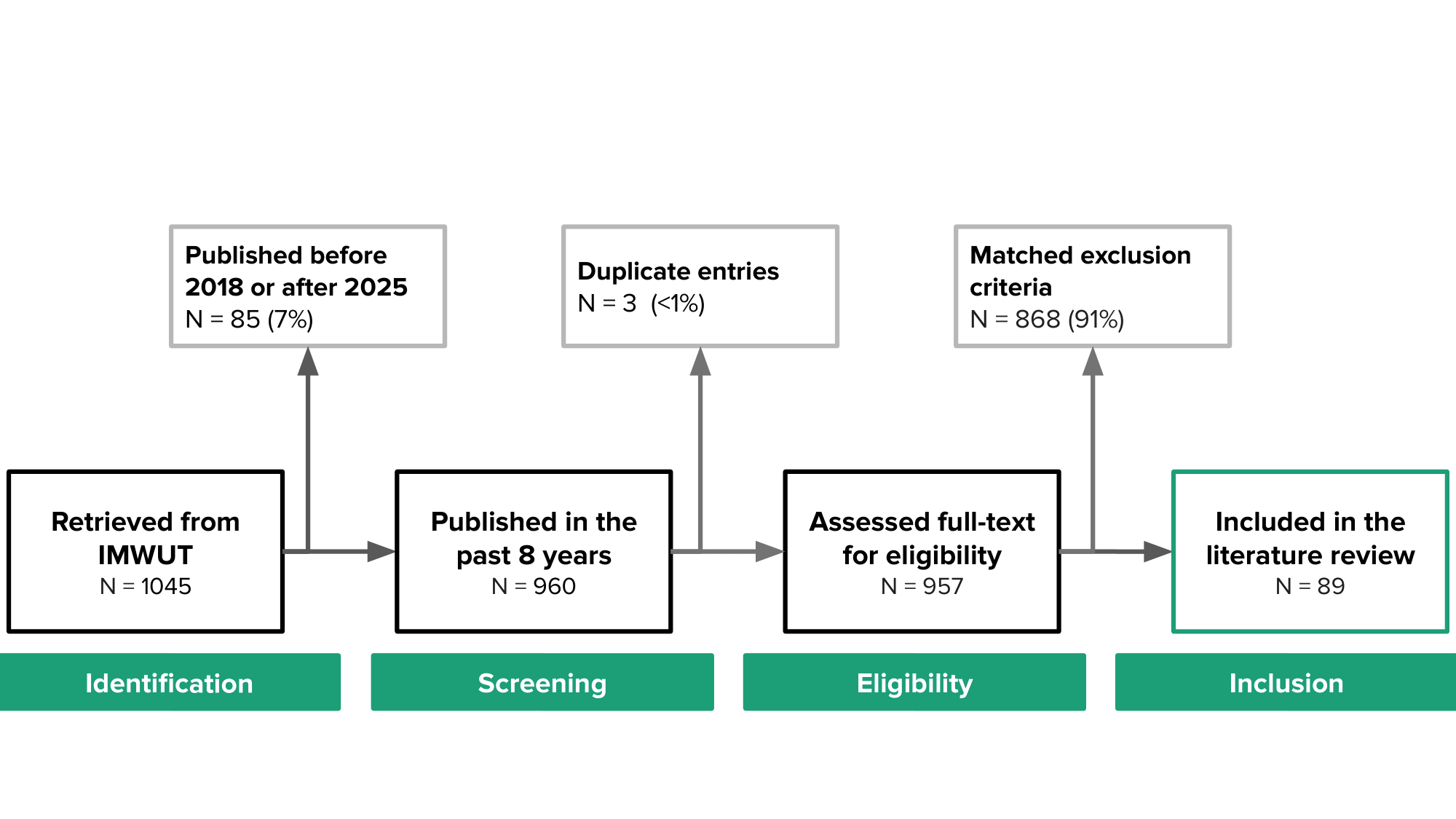}
  \caption{\textbf{PRISMA flow diagram for paper inclusion}. Out of the \rev{957} papers retrieved by our query after the screening, only \rev{9\% ($N=89$)} did not check any exclusion criterion and thus were included in the literature synthesis.\label{fig:prisma}}
\end{figure}

\noindent\textbf{Eligibility Assessment.} To further validate our query, we manually inspected all publications from the latest available IMWUT proceedings \rev{at the time of manuscript preparation} (Volume 6, Issue 4, published in January 2023) ($N=56$) to identify eligible papers for inclusion (see inclusion and exclusion criteria below). In total, we identified seven relevant publications, all of which were also returned by our query. This process was irrelevant to our final paper retrieval (pictured in Figure~\ref{fig:prisma}) and served validation purposes only. To ensure the high quality and relevance of the included papers, we defined appropriate exclusion criteria that helped us determine the included papers:
\begin{enumerate}
    \item Papers that do not provide a quantitative assessment of at least one empirical or artifact contribution in mobile and wearable technologies (UBI);
    \item Papers that do not include a quantitative assessment of bias or performance discrepancy in their evaluation with regard to sensitive attributes, \rev{specifically age, gender, race, country of origin, language, disability or disease status, marital status, religion, employment status, physiology (i.e., weight, height), socioeconomic status and sexual orientation} (FAIR);
    \item Papers that discuss different domains, such as NLP or computer vision, without incorporating a ubiquitous component (DOM);
    \item Papers that refer to bias in a different context, such as the bias-variance trade-off or the bias parameter in neural networks (CON). 
\end{enumerate}

\noindent\textbf{Inclusion \& PRISMA Statement.} Figure~\ref{fig:prisma} shows the Preferred Reporting Items for Systematic Reviews and Meta-Analyses (PRISMA) \cite{moher2009preferred} flow diagram. Specifically, the sequential execution of the steps above led to our review's included papers. Overall, we screened \rev{957} papers after date filtering and duplicate elimination. We then excluded \rev{868} based on our exclusion criteria. Hence, we included \rev{89} papers in our review.\footnote{To foster reproducibility, upon acceptance, we intend to release the review data and codebooks}

\begin{table}[]
\caption{\textbf{To generalize these results across venues of similar scope, we analyzed recent proceedings \rev{(at the time of writing)} of ACM MobiCom, MobiSys, and SenSys, and IEEE Trans. of Mobile Computing and Pervasive Computing, and found no deviation from our primary result.} Out of the 668 total papers published in 2022 in these venues, 376 were retrieved by our query, and only 27 complied with our screening eligibility criteria.}
\label{tab:venues}
\begin{tabular}{llll}
\rowcolor[HTML]{1C9E78} 
{\color[HTML]{FFFFFF} \textbf{Venue}} & {\color[HTML]{FFFFFF} \textbf{\#Published (2022)}} & {\color[HTML]{FFFFFF} \textbf{\#Retrieved (\% total)}} & {\color[HTML]{FFFFFF} \textbf{\#Included (\% total)}} \\
{ACM IMWUT}           & {154}                          & {131 (85\%)}                       & {15 (10\%)}                        \\
\rowcolor[HTML]{F3F3F3} 
MobiCom                               & 56                                                 & 25 (45\%)                                              & 0 (0\%)                                               \\
MobiSys                               & 38                                                 & 18 (47\%)                                              & 3 (8\%)                                               \\
\rowcolor[HTML]{F3F3F3} 
SenSys                                & 52                                                 & 8 (15\%)                                               & 2 (4\%)                                               \\
IEEE Pervasive                        & 49                                                 & 15 (31\%)                                              & 2 (4\%)                                               \\
\rowcolor[HTML]{F3F3F3} 
IEEE Trans. Mob. Comp.              & 319                                                & 179 (56\%)                                             & 5 (2\%)                                               \\
{\textbf{Total}}           & {\textit{668}}                & {\textit{376 (56\%)}}             & {\textit{27 (4\%)}}                                  
\end{tabular}
\end{table}
To assess consistency across venues, we reviewed publications from relevant mobile, wearable, and ubiquitous computing conferences and journals from 2022. Specifically, we reviewed ACM SIGMOBILE-sponsored events, including the International Conference on Mobile Computing and Networking (MobiCom, $N_{2022}=56$), the International Conference on Mobile Systems, Applications, and Services (MobiSys, $N_{2022}=38$), and the ACM Conference on Embedded Networked Sensor Systems (SenSys, $N_{2022}=52$), and relevant IEEE journals, including IEEE Pervasive Computing (IEEE Pervasive, $N_{2022}=49$) and IEEE Transactions on Mobile Computing (IEEE Trans. Mob. Comp., $N_{2022}=319$). Following the same methodology, out of 245 retrieved papers, 12 papers complied with our screening and eligibility criteria: none from MobiCom (0\% inclusion rate), 3 from MobiSys (8\%), 2 from SenSys (4\%), 2 from IEEE Pervasive (4\%), and 5 from IEEE Trans. Mob. Comp. (2\%). Table~\ref{tab:venues} summarizes our analysis.

Overall, while we acknowledge the limitations of considering a single publication for this review, these findings provided an indication of the generalizability of our results across mobile, wearable, and ubiquitous computing publication venues.

\subsection{Positionality Statement\label{positionality}}
Understanding researcher positionality is essential to shed light on our perspective on data collection and analysis~\cite{frluckaj2022gender, havens2020situated}. We situate this review paper in a Western country in the 21\textsuperscript{st} century, writing as authors who primarily work as academic and industry researchers in the technology sector. We identify as two female and four male authors, and our shared backgrounds include mobile, wearable, interactive and ubiquitous computing, human-computer interaction, and machine learning and artificial intelligence. 

\section{Findings: How bias creeps in} 
Our perception of the world is shaped by the way our senses and cognitive processes interpret the information around us. Similarly, personal devices rely on sensory input to make decisions. But here's the catch: these inputs and the algorithms that rely on them are not free from biases (\rev{Figure}~\ref{fig:examples}). 

In summary, of the \rev{957} papers retrieved, only 9\% \rev{($N_{included}=89$)} were included in the review, representing just 5\% of all IMWUT publications from 2018 to \rev{2025}. This underscores the timeliness and necessity of this work. Health emerged as the most prevalent domain, comprising \rev{almost} one-quarter of the included papers. This was followed by \rev{Human-Activity Recognition (20\%), Behavioral Sensing \& Emotion (15\%), Sound, Voice \& Hearing (13\%) and Privacy \& Security (12\%). Less common categories included Mobility \& Navigation (6\%), Motion, Gaze, Gesture \& Touch (5\%),  and Cognition \& Attention (5\%). The least represented domain was Miscellaneous, accounting for 1\% of the included papers.} 

\subsection{Are Personal Devices Susceptible to Biases?}
Several studies have revealed biases in acoustic signals tied to our body's unique characteristics. In respiratory monitoring, higher errors have been reported for users with weaker chest motions and smaller body size \cite{10.1145/3517253}. \rev{Gender-related disparities have likewise been observed, where males’ generally higher respiratory intensity produces stronger abnormal respiratory audio features, while the under-representation of female samples during training limits generalization performance for female users \cite{gong2025sputumlocator}.} Similar effects have been reported in breathing monitoring more broadly \cite{10.1145/3534595}, due to variations of physiological structures (i.e., women's smaller reflective surfaces encode less information).
\rev{In speech and audio processing, gender-related biases have been observed in speech enhancement, where higher vocal frequencies in female speech negatively affect enhancement performance \cite{wang2025acccall}. Complementary findings in speech recovery further indicate a slight performance advantage for male speech, attributed to more consistent acoustic characteristics and stronger low-frequency components \cite{wang2025mmhse}. Race-related biases have also been identified in speech enhancement systems, where hairstyle-associated physical characteristics (e.g., African dreadlocks obstructing the acoustic gap created by auxiliary spacers) resulted in lower signal quality metrics \cite{duan2024earse}. Beyond speech and respiration, biases have been reported in other acoustic sensing and interaction systems.} In acoustic heartbeat monitoring, individuals with higher BMI experienced reduced detection performance due to thicker thoracic musculature leading to weak heartbeat signals \cite{10.1145/3550293}. \rev{Similarly, acoustic interaction systems have shown degraded performance for children under 15, largely due to height constraints below supported operating ranges, as well as for older adults, whose age-related hearing decline increased output energy error, highlighting accessibility limitations tied to age and stature \cite{zhou2024visar}. Finally, prior work on acoustic key generation using bone-conducted vibrations has highlighted age-related physiological differences, showing that variations in bone density affect vibration transmission patterns \cite{wang2025s2pair}.}

\rev{Biases have also been reported in radio-frequency–based sensing modalities, including WiFi, Ultra-wideband (UWB), and millimeter-wave radar systems. In contactless sleep monitoring using UWB signals, sleep apnea severity systematically affects performance \cite{li2024hypnos}, as signals from individuals with more frequent apnea events provide more informative disturbances that aid sleep-stage prediction, while milder cases are more prone to misclassification. RF-based respiratory monitoring further reports reduced accuracy for larger body types, where increased body size may exceed predefined candidate regions, limiting the capture of respiratory motion \cite{wang2024rf}. In mmWave radar–based gait recognition, children’s more variable and playful walking patterns lead to increased gait instability and reduced recognition performance \cite{wang2024rdgait}, while wearable mmWave head-gesture sensing is affected by morphological factors such as long hair (a common proxy for gender), which introduces signal noise due to random motion \cite{yang2023headar}. Finally, WiFi-based body composition sensing shows higher errors for larger individuals due to increased signal attenuation and noise sensitivity, with sex-related differences in abdominal physiology further influencing signal attenuation patterns \cite{pan2024perfat}.}

Biases in computer vision models have also been encountered. For instance, video biases related to physiology, such as height, and appearance, such as hair or head covering (potentially proxies for gender and religion) \cite{10.1145/3161198} or race biases in non-contact photoplethysmography (PPG) measurements \cite{nowara2020meta}. \rev{Vision-based physiological sensing has further revealed performance disparities linked to body morphology and skin-related attributes: infrared vein-based dehydration detection exhibits reduced accuracy for heavier individuals due to diminished vein visibility \cite{islam2025h2opulse}, while wearable optical spectroscopy shows accuracy variations across skin tones, largely attributed to under-representation of certain tones in training data \cite{hamid2025dermaglow}. Thermal imaging–based respiration sensing similarly reports slightly higher errors for participants with overweight BMI, though remaining largely BMI-agnostic compared to smartwatch-based baselines \cite{adhikary2024jouleseye}.}
Some of these biases are easier to distinguish as demographic information is integrated into the data. For example, gender may be inferred from video or speech signals. However, sensor signals ---the most prevalent data modality used by the community--- are more challenging but equally susceptible to data biases.

Biases in sensor signals can be attributed to measurement inaccuracies or concept drift within signal patterns. 
For instance, research has revealed racial biases in pulse oximeters \cite{fawzy2022racial} and PPG sensors for heart rate measurement \cite{bent2020investigating}, while gender biases have been linked to variations in electrocardiogram (ECG) quality \cite{xue2014can}.
Beyond signal acquisition, some studies have shown that the accuracy of certain personal device functions varies with demographic characteristics. Gait detection through accelerometer and gyroscope measurements may be less accurate for elderly users due to age-related changes in gait patterns \cite{10.1145/3351281}. \rev{Similarly, \citet{li2023neuralgait} report performance variations in Inertial Measurement Unit (IMU)-based neurological disease screening across sex and age groups, reflecting sex-specific gait biomechanics and age-related gait changes, with reduced sensitivity for younger subjects due to limited training data and for older adults in stroke and traumatic brain injury screening. IMU- and sEMG-based knee osteoarthritis gait re-training also exhibits higher errors for patients with more severe disease, overweight BMI, and male users, attributed to altered gait patterns, increased body-induced signal noise, and greater muscle loading, as well as limited data availability for severe cases \cite{yang2024kneeguard}. Physiological sensing accuracy is further modulated by demographic characteristics. Sleep staging from ballistocardiogram signals is affected by age-related attenuation of sleep spindles, higher BMI–associated sleep apnea prevalence, and documented gender- and race-related differences in sleep architecture \cite{li2024sleepnetzero}. In cardiovascular monitoring, cuffless wrist-worn blood pressure sensing reports reduced signal quality for overweight individuals due to increased tissue depth over the radial artery, as well as variability linked to hypertension-related vascular stiffness \cite{li2025practicalbp}.} 
In addition, speech recognition based on accelerometers can be substantially influenced by sex-related physiological variations in individuals' voices (e.g., women tend to have thinner and higher-pitched voices), particularly if these individuals are under-represented in the training data) \cite{10.1145/3494969,10.1145/3550338}. IMU and proximity sensors showed performance discrepancies in eating activity detection for users with a large BMI, attributed to factors like changes in the distance of the proximity sensor from the neck \cite{10.1145/3397313}. 
\rev{Comparable body-morphology–related effects have also been observed in pressure-based sensing, where human shape estimation from pressure image sequences achieves higher accuracy for subjects closer to average body weight, while performance degrades for lighter or heavier individuals due to distributional shifts in body shape \cite{wu2024seeing}.}
As models become larger and their training objectives more universal, they learn to generalize to more tasks, including sensitive attributes' inference.
Even sensors as seemingly inconspicuous as those used for heart rate and step tracking can be used to predict fine-grained outcomes relevant to health, fitness, and demographic characteristics \cite{spathis2021self}. 
However, the challenge with biases in personal devices is that they are often not straightforward to detect. They may remain blended in the background within the signal data or, worse, propagate into high-level inferences, affecting the assessments of health and well-being.

 \begin{figure}[]
  \centering
  \includegraphics[width=.95\linewidth]{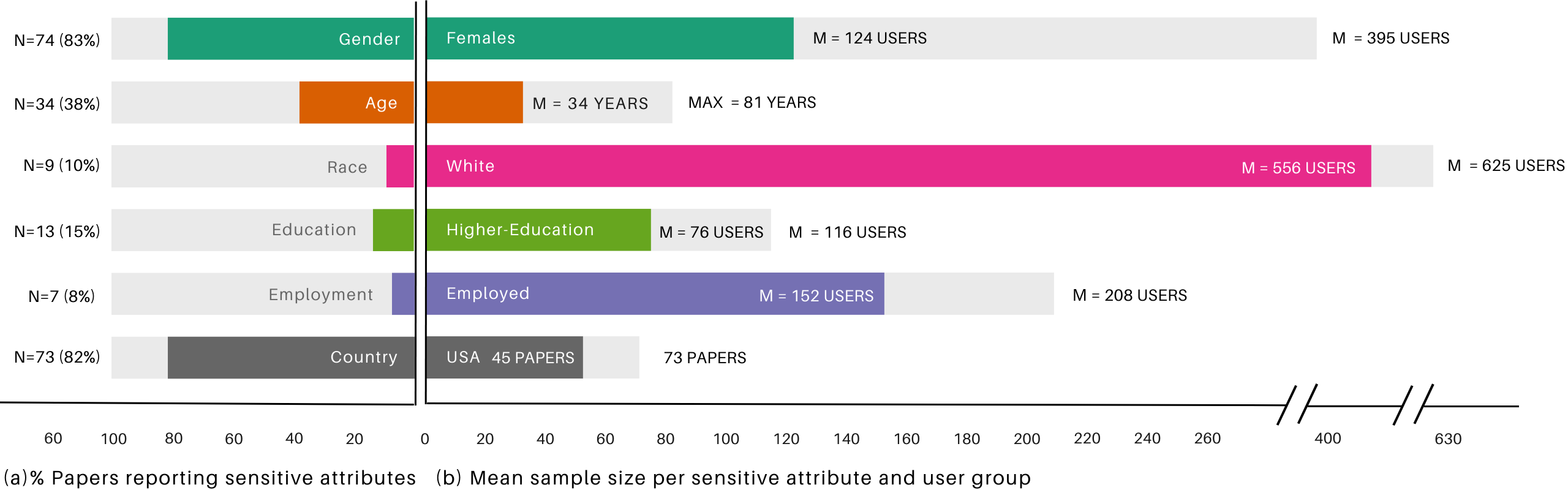}
  \caption{\textbf{Analysis of sensitive attributes and data size}. The bar plots show the percentage of papers reporting certain sensitive attributes (left) and the (mean) sample size in the subset of papers reporting that attribute (right). Sample demographics reporting is not standardized and frequently incomplete, with race, employment status, and education being the least reported sensitive attributes.\label{fig:weird}} 
  \vspace{-0.15in}
\end{figure}
These hidden biases persist due to disparities in demographic representation within commonly employed datasets, alongside the under-reporting of demographic variables.
For example, the recruitment of predominantly WEIRD samples is a pressing concern within the mobile and wearable community. Figure~\ref{fig:weird} shows an analysis of sensitive attributes across papers published in IMWUT.
Gender is the most reported sensitive attribute with the mean sample size being \rev{395} users, and the mean number of females being \rev{124 (31\%)}. Evidently, while the community has taken steps in the right direction, there is still plenty of room for improvement.
Out of the papers for which the participants' country is reported or can be inferred, almost \rev{6} out of 10 engaged with US samples. 
Indicatively, age is only reported in \rev{less than} 4 out of 10 papers, with the mean sample age being \rev{34} years old. For comparison, the maximum mean age reported is \rev{81} years. In a world that is rapidly aging, personal devices are being predominantly developed and tested on young populations. 
Concerning education, \rev{less than} 2 in 10 papers reported relevant information, with a median sample size of \rev{116} users, \rev{76} of which were college-educated \rev{(66\%)}. This is perhaps not surprising, as in the early stages of personal device research, participant recruitment frequently occurs within universities. 
Also, a mere 1 in 10 papers mentioned the participants' race. Among those papers, the sample size is an average of \rev{625} users, with \rev{556} of them being White \rev{(89\%)}. These data highlight not only that race is often overlooked but also that non-White populations are underrepresented in the utilized datasets. This poses a real risk that models underperforming for non-White users may go unnoticed. For instance, prior work could not assess the impact of skin tone on AFib detection due to the majority (88.7\%) of participants being White \cite{10.1145/3463503}.

\subsection{How do we Mitigate Biases?}
It is often infeasible to eliminate all sources of unfairness. Yet, the goal is to surface and mitigate biases as much as possible through fairness enhancement mechanisms. 

Within the mobile, wearable, and ubiquitous computing research, preliminary pre-processing mechanisms for fairness enhancement included fair data representation. For instance, \citet{10.1145/3328927} included both healthy and non-healthy subjects in their dataset for respiratory rate monitoring. Similarly, \citet{10.1145/3534585} employed a fairness-aware client selection mechanism for federated learning to ensure equal representation for subjects with worse connectivity.\footnote{While Internet connectivity is not a sensitive attribute per se, it has been linked with socioeconomic status, race, nationality, gender, and age, all of which are sensitive attributes \cite{united2021almost}.} 
\citet{10.1145/3494969} performed data balancing managing to narrow the impact of gender voice differences on their speech recognition model. \rev{Along similar lines, \citet{wang2025acccall} explicitly recruited a gender-balanced evaluation cohort to assess and reduce disparities in accelerometer-based speech recognition.}
Similarly, a strand of work explored data splitting, conditioned on the sensitive attribute (gender, age, BMI, skin tone, country, and health condition) to enable model personalization \cite{10.1145/3369820,10.1145/3351246,10.1145/3517249,10.1145/3494969,10.1145/3351281}. More advanced mechanisms suggest modifying feature representations. For example, \citet{10.1145/3191774} improved their activity detection model performance by normalizing the window-level features across gender and physiology. Similarly, in line with prior work~\cite{locatello2019fairness}, \citet{10.1145/3517252} utilized disentangled representations, aiming to isolate activity patterns from redundant noises such as gender, age, and physiological differences.

Regarding in-processing mechanisms, \citet{10.1145/3411835} altered their logistic regression model for Post-traumatic stress disorder (PTSD) screening to include sensitive attributes in its parameters. On a similar note, \citet{10.1145/3397330} devised a multi-task loss function consisting of activity, subject, and gender loss. \rev{\citet{li2024hypnos} similarly leveraged multi-task learning, where signals from subjects with more severe sleep apnea provided informative auxiliary supervision that helped correct sleep-stage predictions. Along the same line, \citet{10.1145/3569483} have shown that partially personalized country- or continent-specific models outperform country-agnostic models in depression screening, highlighting the role of group-conditioned in-processing strategies in mitigating performance disparities across geographic populations.} Finally, in quantifying the causal effect of individual mobility on health status, \citet{10.1145/3494990} considered correlated sensitive attributes, such as age and socioeconomic status, as confounding variables in their causal model.
However, they noted that due to dataset privacy constraints in reporting demographic variables, potential unobserved confounding variables might have been missed, highlighting the conflict between fairness and privacy \cite{chang2021privacy}. \rev{At the system level, \citet{djebrouni2024bias} proposed a bias-constrained federated learning aggregation approach that enforces a predefined fairness objective by constraining bias below a threshold while maintaining accuracy.}

\rev{Finally, post-processing mechanisms aim to mitigate bias by correcting measurements or adapting model outputs after model training. For instance, \citet{hamid2025dermaglow} applied bias-aware calibration by leveraging skin-tone features and device identifiers to produce corrected oxygen saturation estimates, reducing skin-tone–dependent errors in pulse oximetry. Similarly, \citet{wang2025s2pair} employed per-person bone-channel estimation models to compensate for age-related differences in bone density. \citet{li2025practicalbp} combined single-point calibration with feature-wise attention to assess the importance of each feature for different subjects and adapt blood pressure estimation to individual physiological patterns.}

Despite the significant efforts of the aforementioned pioneering works in the community, 3 out of 4 included papers did not report any fairness enhancement mechanism, regardless of the presence of bias in their models. This is partly due to a lack of consideration for fairness-related harms, but it is also connected with the nature of several personal device works: artifact contributions, proof of concept, and early-stage technology development, where performance is prioritized.

\begin{table}[htb!]
\caption{\textbf{Fourteen Recommendations for ``Fairness by design'' in personal devices.} Guidelines for researchers for performing fairness assessments given challenges specific to personal devices, accompanied by examples.}\label{fig:recommendations}
\resizebox{\textwidth}{!}{%
\begin{tabular}{lp{2.3in}p{3.4in}p{2in}}                                            
\multicolumn{1}{c}{{\color[HTML]{1C9E77} \textbf{\#}}} & \multicolumn{1}{c}{{\color[HTML]{1C9E77} \textbf{Generic Recommendations}}}                                  & \multicolumn{1}{c}{{\color[HTML]{1C9E77} \textbf{Recommendations Tailored to Personal Devices}}}                                                                                                   & \multicolumn{1}{c}{{\color[HTML]{1C9E77} \textbf{Example}}}                                                                        \\
\rowcolor[HTML]{1C9E77} 
\multicolumn{4}{c}{\cellcolor[HTML]{1C9E77}{\color[HTML]{FFFFFF} \textbf{DATA COLLECTION \& ANNOTATION}}}                                                                                                                                                                                                                                                                                                                                                                       \\
\rowcolor[HTML]{EFEFEF} 
1                                                      & Identify biases across demographics during design                                            & Consider diverse physiological characteristics beyond legally protected attributes for fair representation                                                                         & Acoustic sensing exhibits weight bias \cite{10.1145/3550293}                                                                                    \\
2                                                      & Recruit adequate sample sizes to ensure diverse representation of different groups           & Consider contextual variables specific to the individual's environment and lifestyle and address variable engagement levels across demographics in personal devices                & AFib detectors consider abnormal heart rhythms under a single label due to small sample sizes \cite{10.1145/3397313}                                                      \\
\rowcolor[HTML]{EFEFEF} 
3                                                      & Ensure that the ground truth data labeling process involves a diverse group of annotators    & Train annotators on implicit biases to cultivate fair representation in human-centric personal device data labeling                                                   & Personal devices datasets are often self-annotated (by WEIRD authors)                                                             \\
4                                                      & Report protected attributes in datasets                                                      & Ensure transparency around sensitive attributes, including proxy features that might introduce bias to predictions, considering the "unreadable" nature of the data                & Confounding variables may be missed due to privacy constraints in reporting protected attributes in HAR datasets \cite{chang2021privacy}                \\
\rowcolor[HTML]{1C9E77} 
\multicolumn{4}{c}{\cellcolor[HTML]{1C9E77}{\color[HTML]{FFFFFF} \textbf{DATA EXPLORATION \& MANIPULATION}}}                                                                                                                                                                                                                                                                                                                                                                    \\
\rowcolor[HTML]{EFEFEF} 
5                                                      & Validate data across groups to detect anomalies                                              & Develop data validation methods to normalize across factors like device placement and user activity levels, mitigating biases introduced by in-the-wild variability     & Sensor outliers can bias certain groups                                                                                            \\
6                                                      & Investigate sources of measurement error                                                     & Examine sources of measurement error from device hardware in model-based estimates to enhance accuracy in downstream tasks                                                        & Sensor drift over time can cause discrepancy between the physical state that is measured and the sensor output                                \\
\rowcolor[HTML]{EFEFEF} 
7                                                      & Be considerate of how data pre-processing methods affect diverse demographics                & Consider the impact of windowing on data representation, design choices, and inconsistent data due to device fit and no wear time                                               & Sampling rate or model pruning and compression can bias on-device ML \cite{toussaint2022tiny}                                                                 \\
\rowcolor[HTML]{1C9E77} 
\multicolumn{4}{c}{\cellcolor[HTML]{1C9E77}{\color[HTML]{FFFFFF} \textbf{MODEL DEVELOPMENT \& EVALUATION}}}                                                                                                                                                                                                                                                                                                                                                                     \\
8                                                      & Consider out-of-the-box fairness enhancement methods to mitigate bias in the data and models & Prioritize lightweight neural architectures and adaptive bias mitigation algorithms, addressing limited software library support for common device ML tasks                                    & Research on bias mitigation for multi-class classification and regression tasks lags behind \cite{agarwal2019fair}                                       \\
\rowcolor[HTML]{EFEFEF} 
9                                                     & Consider indirect notions of fairness                                                        & Address fairness issues arising from device diversity and connectivity                                                                                                             & Client exclusion due to unsupported devices in federated learning \cite{cho2022flame} \\
10                                                     & Assess model performance using varied fairness metrics for intersectional user groups        & Develop fairness metrics suitable for real-time assessment on personal devices and consider the ambiguity in defining fair performance thresholds, especially for regression tasks & Fairness metrics offer only static snapshots, failing to capture real-time fairness dynamics in personal devices                                                                     \\
\rowcolor[HTML]{EFEFEF} 
11                                                     & Generate diverse synthetic data covering various protected attributes and intersections      & Explore timeseries-specific data generation methods to ensure fairness in synthetic personal devices data                                                                         & Personal devices datasets often lack size to enable fairness assessments \cite{yfantidou2023uncovering}                                                           \\
\rowcolor[HTML]{EFEFEF} 

\rowcolor[HTML]{1C9E77} 
\multicolumn{4}{c}{\cellcolor[HTML]{1C9E77}{\color[HTML]{FFFFFF} \textbf{MODEL DEPLOYMENT AND USE}}}                                                                                                                                                                                                                                                                                                                                                                            \\
12                                                     & Monitor real-time performance of deployed models                                             & Identify and adjust for data (sensor) and fairness drift                                                                                                                           & Fall detectors trained on younger populations and deployed on the elderly                                                          \\
\rowcolor[HTML]{EFEFEF} 
13                                                     & Reassess the model using collected data after deployment to ensure continued fairness        & Account for challenges like lack of labels post-deployment by utilizing unsupervised protocols                                                                                     & In fall detection ground truth data are unavailable without user reporting                                                         \\
14                                                     & Provide transparency in model versioning to increase accountability and reproducibility      & Ensure transparency in changes to estimate models, as modifications may impact multiple downstream tasks, affecting user experience if unreported                                  & Changes in algorithms cause drifts in unexpected downstream tasks that may disproportionately affect user groups                  
\end{tabular}%
}
\vspace{-0.2in}
\end{table}

\subsection{How does the Community Capture Alternative Notions of Fairness?}
Previously, we have established that only a small fraction of retrieved works ($\sim$9\%) provide conventional fairness definitions with respect to one or more sensitive attributes. Yet, we believe such definitions do not do full justice to the community's work, which strives for ``fairer'' models, perhaps not across sensitive attributes but differing experimental conditions. In particular, we noticed that, in evaluating new personal device systems, the community aims for generalizable and robust models by performing ablations studies, comparing deployment settings, and personalizing models for users and groups \rev{(67\% of included papers)}. 
\smallskip 

\noindent\textbf{Ablation Studies.} In an ablation study, one or more components of the model are systematically removed or modified, and the performance of the model is evaluated after each change, ensuring its generalizability and robustness. In the included papers, ablation studies take the form of performance evaluation comparisons based on: \textit{user-related}, \textit{device-related}, \textit{environmental}, \textit{experimental}, and \textit{domain-specific components}. In particular \textit{user-related components} include user motion and orientation during data collection in sleep posture monitoring~\cite{10.1145/3397311}, respiratory monitoring~\cite{10.1145/3534595,wang2024rf}, gesture recognition~\cite{10.1145/3432235}, \rev{user identification~\cite{10.1145/3264944,wang2024rdgait,he2025ht}}, \rev{indoor localization \cite{chen2023environment}, }\rev{audio processing and speech generation \cite{wang2025acccall,shah2024stethospeech}}, and heart activity monitoring~\cite{10.1145/3478127}, as well as \rev{physiological characteristics and aesthetics, such as neck circumference}, hair or clothing in fine-grained activity sensing~\cite{10.1145/3517253}, breathing and vital sign monitoring~\cite{10.1145/3534595,10.1145/3550293}, \rev{sleep monitoring \cite{shao2025my}} and user identification~\cite{10.1145/3161198,10.1145/3351273}. 
\textit{Device-related components} include device type, placement, and orientation, sampling rate, and operating system in activity and gaze tracking~\cite{10.1145/3517253,10.1145/3494999}, vital sign monitoring and physiological sensing~\cite{10.1145/3550293,10.1145/3517225}, speech recognition via built-in sensors and speech synthesis \cite{10.1145/3494969,10.1145/3550338}, \rev{audio processing \cite{wang2025acccall}} and \rev{user behavior sensing \cite{10.1145/3448089,wang2024rdgait,xiao2024reading}}. \textit{Environmental components} include ambient noise, light, and temperature that might affect data quality of acoustic~\cite{10.1145/3517253,10.1145/3448113,10.1145/3550293,shah2024stethospeech,chen2023voicecloak} or video~\cite{10.1145/3191772,10.1145/3161164,10.1145/3517225,10.1145/3161198} signals, or random passers-by for human identification \cite{10.1145/3351273}. Regarding experimental setup, few included papers studied the effect of equipment placement (e.g., distance, angle) and characteristics (e.g., range) on the model's robustness in activity sensing~\cite{10.1145/3517253}, \rev{audio processing \cite{wang2025acccall}} and vital sign monitoring using acoustic signals~\cite{10.1145/3534595,10.1145/3550293}. \rev{Experimental setups also include the effect of hyperparameters, such as data size, number of clients in federated learning, or deep learning architectural choices on performance \cite{djebrouni2024bias,xiao2024reading,song2025multi}.} Yet, the choice of components to consider in an ablation study is highly domain-dependent. \textit{Domain-specific components} have no limitations and can range from screen size in scrolling interaction experiments \cite{10.1145/3351255} to food structure in food-related artifact development \cite{10.1145/3550312} \rev{and bed size and mattress firmness in sleep posture recognition \cite{liu2024tagsleep3d}}.
\smallskip

\noindent\textbf{Deployment Setting.} A study's deployment setting, whether it be in-the-lab or in-the-wild, can significantly impact its outcomes. While laboratory settings can provide controlled environments for experimentation, they may not accurately reflect the complexities of the real world in which the applications are deployed. As a result, in-the-wild (or in-situ) studies have emerged as an alternative, focusing on evaluating the situated design experience of mobile, wearable, and ubiquitous computing. In the included papers, perhaps not surprisingly, in-the-lab studies prevail \rev{($\sim54\%$)}, which can be explained by the nature of numerous IMWUT papers presenting artifacts or early-stage work. Nevertheless, \rev{$\sim29\%$} of included papers conduct in-the-wild studies, and a small fraction of papers \rev{($\sim8.5\%$)} report results for both deployments. Interestingly, while 3 out of 10 in-the-lab studies do not identify biases, this number falls to 0.5 for in-the-wild studies. This confirms our intuition that controlled environments can conceal biases that would emerge once a model is deployed in the real world.
\smallskip 

\noindent\textbf{Personalization.} 1 in 5 papers trained separate personalized models for inference \rev{on a single subject \cite{10.1145/3351281,10.1145/3264944,10.1145/3161601,wang2025s2pair,li2025practicalbp,shao2025my} or a group of subjects sharing a common characteristic \cite{yfantidou2023uncovering,10.1145/3569483}}. For instance, \citet{10.1145/3369820} built personalized models for different age groups ``\textit{illustrating differences in communication patterns across age demographics that can impact model performance}''. Similarly, \citet{10.1145/3517249} utilized age-specific models for just-in-time mobile safety help, as ``\textit{different ages in the sample have a significant influence on supportable moment predictions}''. \citet{10.1145/3517225} developed personalized models based on skin tone for camera-based photoplethysmography, as ``\textit{previous work had already highlighted [skin tone and gender] issues with the Plane-Orthogonal-to-Skin [method]}''. \citet{10.1145/3494969} developed gender-specific models for speech recognition, as ``\textit{women’s voice is generally thinner and higher in pitch}'', while \citet{10.1145/3397313} explored BMI-based models for detecting eating activities via a multi-sensor necklace, due to ``\textit{differences in movement patterns while eating, change in the distance of the proximity sensor from the neck, and difference in posture during the eating activity}''. 

Overall, the gains from the interaction between the mobile, wearable, and ubiquitous computing and fairness communities can be bi-directional. On one side, personal device research can learn from other communities' best practices in fairness assessment and reporting. Still, the latter can build on this community's approach of rigorous validation, model personalization, and striving for robustness and generalizability through in-the-wild deployments and comprehensive ablation studies.

\section{The way forward: guidelines for making personal devices fair}
Borrowing from ``Privacy by design'' \cite{cavoukian2009privacy}, we propose a ``Fairness by design'' equivalent, requiring AI developers and researchers to consider data and model fairness from the very beginning of any system design.
This proactive approach prioritizes fairness as a core value in the development of personal devices. To guide the mobile and wearable computing community, we offer fourteen recommendations for seamlessly integrating fairness throughout the ML pipeline (Figure~\ref{fig:recommendations}). These guidelines span across data collection and annotation, data exploration and manipulation, model development and evaluation, and finally, model deployment and use.
\smallskip

\noindent\textbf{Data Collection \& Annotation.} For data collection, researchers should \textit{identify the types of biases and harms} (\#1) relevant to their task (e.g., quality-of-service, allocation, and erasure harms \cite{crawford2017trouble}). For example, in an AFib detection application, quality-of-service harms could occur if the model had a substantially different performance across ages. In contrast, allocation harms could occur if such difference led to one group unfairly receiving worse care. It is also important to consider group demographics (e.g., historically marginalized groups) relevant to a particular deployment setting that might be harmed. In personal devices, besides legally mandated protected attributes, users' physiological characteristics like height, weight, or health conditions must also be considered. For example, acoustic sensing for vital sign monitoring applications can be less accurate for overweight and obese people \cite{10.1145/3550293}. While comprehensive coverage of sensitive attributes might not be guaranteed, we highly recommend consulting domain experts or conducting feasibility studies in the target application domain.

Ideally, researchers should \textit{collect adequate and diverse sample sizes} (\#2) tailored to the application's needs and target audience to ensure the generalizability of the prediction task. For example, AFib detection systems have shown poor performance in people with abnormal heart rhythms due to limited data diversity in terms of subjects with different types of heart conditions \cite{10.1145/3397313}. Best practices in publishing ubiquitous datasets with protected attributes include large-scale datasets such as MyHeart Counts \cite{hershman2019physical}, GLOBEM \cite{xu2023globem}, and MIMIC \cite{johnson2016mimic} datasets, as well as smaller datasets such as LifeSnaps \cite{yfantidou2022lifesnaps}, MotionSense \cite{Malekzadeh:2019:MSD:3302505.3310068}, and ExtraSensory \cite{vaizman2017recognizing}. However, while acknowledging the impracticality of recruiting large samples in feasibility studies, researchers should prioritize diversity in terms of relevant protected attributes to mitigate biases in later development stages. Interestingly, factors such as geographic location, daily routines, and cultural practices can significantly impact the data generated by personal devices. 
Additionally, achieving representative samples in wearable technology may be challenging due to variable engagement levels across demographics. 
Yet, diverse sampling must balance representing the target audience with maintaining a learnable data distribution to avoid noise and outliers. 

For data annotation, researchers should aim for \textit{representation from diverse annotators} (\#3) and offer implicit bias training. This holds particularly true in scenarios such as personal device applications where labels characterize human subjects as annotators' beliefs and identities can lead to biased label assignment \cite{sap-etal-2022-annotators}.

Additionally, researchers should \textit{supplement published datasets with protected attributes} (\#4) to facilitate fairness analyses. This requires anonymization and secure sharing through designated platforms such as PhysioNet. In signal data, proxy features may introduce bias to downstream tasks \cite{spathis2021self}, complicating debugging due to the data's obscured nature. For example, \cite{chang2021privacy} emphasized the trade-off between fairness and privacy, noting how dataset privacy constraints may obscure potential unobserved confounding variables. In such scenarios, researchers should consider ``fairness in unawareness'' solutions, wherein disparity is assessed even when the protected class remains unobserved \cite{chen2019fairness}.
At the same time, the mobile and wearable research community itself could implement a mandatory data statement policy, requiring authors to report sensitive attributes concerning their participant samples in line with recent quests for data excellence~\cite{sambasivan2021everyone}.
\smallskip

\noindent\textbf{Data Exploration \& Manipulation.} 
For data exploration, data validation methods should be applied across sensitive attributes to facilitate fairness. For example, \textit{inspecting outliers or missing values} (\#5) that systematically fall into particular demographic groups. Unlike controlled laboratory settings, data gathered in-the-wild often contain noise and uncertainty due to factors such as sensor inaccuracies, environmental fluctuations, and user variability. They can also be affected by factors like device placement and user activity levels.

Similarly, \textit{measurement errors} (\#6), namely input data containing inaccuracies, for example due to device malfunctions or variations, can affect fairness in downstream tasks. For example, data from wearables might exhibit anomalies due to device fit issues on smaller wrists typically found in women or children. Yet, in such devices, reliance on model-based estimates for various well-being metrics means that these errors can propagate to subsequent tasks. If the original model has not been validated across different groups, it can adversely affect any applications utilizing such data. In contrast, other domains, like computer vision, are less affected by hardware variations; for instance, the camera used typically has minimal impact on model outcomes. For this purpose, visualization tools (e.g., What-If Tool, FairLens, Tensorflow's Fairness Indicators) can help surface potential data anomalies before they creep into models.

In data manipulation, researchers must \textit{scrutinize pre-processing decisions} (\#7). Unlike fields with out-of-the-box datasets, device data often necessitates additional windowing before predictive modeling. For instance, applying a sliding window method can yield numerous samples from a single user's sensor signal, potentially leading to an overabundance of training data from a limited user pool. As a result, the model may not learn generalizable patterns. Similarly, design choices, such as sampling rate or model pruning and compression, can lead to bias propagation on on-device ML workflows \cite{toussaint2022tiny}. Illustrative examples of good practice include data balancing techniques to reduce the impact of sex-related voice differences in speech recognition models \cite{10.1145/3494969}, or data splitting, conditioned on sensitive attributes to enable model personalization \cite{10.1145/3494969,10.1145/3351281}. More advanced mechanisms suggest normalizing window-level features across demographics and physiology, or utilizing disentangled representations, aiming to isolate activity patterns from redundant noises such as demographic differences.
\smallskip

\noindent\textbf{Model Training \& Evaluation.} 
In model training, while we have witnessed a remarkable consolidation of deep learning models and architectures in personal devices, such as Convolutional Neural Networks and Transformers, fairness-preserving mechanisms remain unexplored. However, there is a plethora of fairness toolkits, such as FairLearn, AIF360 and Aequitas, which provide diverse pre-processing, in-processing, and post-processing \textit{bias mitigation algorithms out-of-the-box} (\#8) --- some applicable regardless of input data types. However, many libraries fail to provide sufficient support for time-series tasks, particularly multi-class classification and regression, which are common in personal device research. Similarly, adaptive bias mitigation algorithms capable of dynamically adjusting to changes in user contexts and device conditions are also lacking. As examples of good practice, in classification tasks, fairness-preserving mechanisms include the incorporation of sensitive attributes as model parameters in Post-traumatic stress disorder (PTSD) screening \cite{10.1145/3411835}, or the adaptation of the loss function to incorporate activity, subject, and gender loss in human-activity recognition \cite{10.1145/3397330}. Other mechanisms also include adversarial debiasing (i.e., training a model to simultaneously classify inputs accurately while minimizing the ability of an adversary to infer sensitive attributes), and discrimination-aware regularization (i.e., penalizing the model for making decisions based on sensitive attributes)
Also, particularly in the context of healthcare, it is important to consider the fairness-utility trade-off, where optimally fair algorithms may result in poorer decisions across groups \cite{10.1145/3097983.3098095}.
More broadly, as personal devices demand lightweight neural architectures, there have been efforts to find neural networks that strike a balance between fairness and accuracy, all while ensuring compliance with hardware specifications \cite{sheng2022larger}.

Similarly, 
we should also \textit{consider indirect notions of fairness} (\#9) during training. For example, within the paradigm of federated learning, the resource allocation of participating devices may also reflect the demographic and socio-economic information of owners, which makes the exclusion of such clients unfair in terms of participation. Cheaper devices cannot support the execution of large models and are either excluded or dropped together with their unique data \cite{cho2022flame}. 
For example, prior research \cite{10.1145/3534585} demonstrates good practices in fair resource allocation in wearable devices of inferior networking conditions (linked with socioeconomic status, race, nationality, gender, and age). 
Regardless of the training paradigm, one should consider additional constraints, such as device diversity or connectivity.

For model evaluation, assessing fairness in personal devices presents a dual challenge: the definition of a fair performance threshold is ambiguous, and fairness evaluation in regression or multi-class classification is understudied, particularly in online scenarios. Establishing a performance threshold is complex, with statistical tests like the Student's t-test facing limitations in cross-validation. 
McNemar's Test and $5\times2$ cross-validation and its refinements are better alternatives proposed \cite{dietterich1998approximate}. 
In classification scenarios, fairness metrics offer alternatives, but the absence of a universal fairness definition hinders their applicability. In regression, fairness definitions are rare and employ measures like statistical parity and bounded group loss~\cite{agarwal2019fair}. 
Comparing outcome distributions through Kullback–Leibler divergence is also used. However, mainstream fairness libraries like AIF360 and FairLearn lack regression-specific metrics.
At any rate, single evaluation metrics struggle to reflect the quality of ML models over time. As such, \textit{monitoring and reporting lightweight fairness metrics} (\#10) suitable for real-time assessment on ubiquitous devices, considering constraints such as limited computational resources but also drift and online learning should become standard practice. Fairness trees accompanied by domain expertise can help researchers choose appropriate metrics. 
A practical example can be drawn from \cite{yfantidou2023uncovering}, where biases are quantified across different stages of the ML lifecycle.

At the same time, we need to account for intersectional fairness, i.e., designing algorithms to account for different social identities that can intersect and impact a person's experiences and outcomes. 
However, we acknowledge that sometimes it is not feasible to collect data from representative demographics to enable such comparisons, especially for smaller pilot studies. In these cases, researchers can leverage advances in time-series- or sensor-specific generative models to \textit{synthesize data covering multiple protected attributes} (\#11) and potential intersections \cite{dahmen2019synsys,chaudhari2022fairgen}. However, it is essential to acknowledge that these models are currently experimental, and rigorous fairness assessments have not yet been performed.
\smallskip

\noindent\textbf{Model Deployment \& Use.}
For model deployment, we should \textit{monitor models' performance in real-time} (\#12) ensuring fair predictions despite shifts in input data and demographics. For instance, in fall detection systems, models are typically trained on simulated falls by young individuals, yet are used to detect real falls in elderly.  Sensor drift can also introduce imperceptible changes in sensors' outputs over time due to factors like temperature, humidity, and hardware aging. These changes can lead to measurement inaccuracies disproportionately affecting certain regions or demographics.

The impact of data drift can be mitigated during deployment through \textit{ongoing model reassessments and refinements} (\#13) tailored to the target population. However, addressing this in personal devices presents challenges, particularly in the absence of post-deployment labels. For example, in fall detection systems, obtaining ground truth data on falls relies on user self-reporting. To address this issue, unsupervised online domain adaptation has been proposed to prevent models from underperforming when applied to new users \cite{10.1145/3432230}.
Additionally, to ensure that model behavior is generalizable for a larger group of users, researchers can also employ newly developed cross-dataset benchmarks for longitudinal models \cite{xu2023globem}. 

During model use, researchers should consider that model refinements can disproportionally affect certain groups. For example, changes in heart-rate prediction models, even if beneficial for minority groups, can propagate to resting heart rate estimations, affecting the user experience. 
Thus, researchers should \textit{prioritize transparency in model versioning} (\#14) and accompany each version with reports on performance and fairness. 
This is particularly relevant for personal devices, as they rely on several ML estimates to determine what is presented to users as their body's ground truth.

\section{Conclusions}
In our technology-laden lives, the pervasive integration of personal devices, diligently monitoring our health and behavior, poses a critical question: Do these digital companions play fair? This article contends that despite their high-stakes applications, personal devices harbor biases, a concern that has been conspicuously overlooked. Drawing attention to the dearth of fairness assessments and the unique characteristics of personal device data, this work advocates for a proactive stance termed ``Fairness by design''. As such, it
moves beyond traditional accuracy evaluations to ensure personal devices not only deliver accurate insights but also adhere to principles of equity and justice for all users.
\begin{acks}
This project has received funding from the European Union’s Horizon 2020 research and innovation programme under the Marie Skłodowska-Curie grant agreement No 813162. The content of this paper reflects only the authors' view, and the Agency and the Commission are not responsible for any use that may be made of the information it contains. The authors acknowledge the use of generative AI tools for assistance with language editing and stylistic refinement. All scientific content, analysis, interpretations, and conclusions are the authors’ own, and the authors take full responsibility for the content of the manuscript.
\end{acks}

\bibliographystyle{ACM-Reference-Format}
\bibliography{main}

\end{document}